\documentclass[]{mn2e}

\onecolumn

\usepackage{epsf}

\newcommand{\source}{\vec{s}}
\newcommand{\data}{{\vec d}}
\newcommand{\pot}{{\vec \psi}}
\newcommand{\dpot}{{\delta \vec \psi}}
\newcommand{\LO}{{\bf L}}
\newcommand{\BO}{{\bf B}}
\newcommand{\OO}{{\bf M}}
\newcommand{\OOT}{{\bf M}^T}
\newcommand{\DS}{{\bf D}_{s}(\source_{\rm p})}
\newcommand{\DP}{{\bf D}_{\psi}}
\newcommand{\CO}{{\bf{L}_{\rm c}(\pot_{\rm p}, {\vec s}_{\rm p})}}
\newcommand{\MO}{{\bf{M}_{\rm c}}}
\newcommand{\MOT}{{\bf{M}^T_{\rm c}}}
\newcommand{\vd}{\vec r}

\newcommand{\sn}{\smallskip\noindent}

\begin{document}

\title[Gravitational Imaging of CDM Substructure]
{Gravitational Imaging of CDM Substructure}

\author[Koopmans]{L.V.E.~Koopmans\\~\\Kapteyn Astronomical Institute,
P.O.Box 800, 9700\,AV Groningen, The Netherlands}

\pubyear{2005}

\label{firstpage}

\maketitle

\begin{abstract}
We propose the novel method of ``gravitational imaging'' to detect and
quantify luminous and dark-matter substructure in gravitational-lens
galaxies. The method utilizes highly-magnified Einstein rings and arcs
as sensitive probes of small perturbations in the lens potential (due
to the presence of mass substructure), reconstructing the
gravitational lens potential non-parametrically.  Numerical
simulations show that the implemented algorithm can reconstruct the
smooth mass distribution of a typical lens galaxy -- exhibiting
reasonable signal-to-noise Einstein rings -- as well as compact
substructure with masses as low as M$_{\rm
sub}$$\sim$$10^{-3}$\,M$_{\rm lens}$, if present. ``Gravitational
imaging'' of pure dark-matter substructure around massive galaxies can
provide a new window on the standard cold-dark-matter paradigm, using
very different physics than ground-based direct-detection experiments,
and probe the hierarchical structure-formation model which predicts
this substructure to exist in great abundance.
\end{abstract}


\vspace{-0.3cm}
\section{Introduction}

Arcsecond-scale strong gravitational lens systems (i.e.\ those with
multiple lensed images) provide a wealth of information about
cosmology, the lensed source and the lens galaxies themselves and
considerable progress has been made in all three fields in the last
twenty-five years (see e.g.\ Kochanek, Schneider \& Wambsganss
2004). However, the accuracy in many applications (e.g. measuring
H$_0$; Refsdal 1964) is set by the poor understanding of the mass
distribution of the lens potential, limiting at the same time the use
of strong lenses as probes of galaxy structure (e.g.\ Kochanek 1991).

\sn It is therefore critical to exploit all available information to
constrain the lens potential, for example by including non-lensing
data (e.g.\ stellar dynamics; Koopmans \& Treu 2002) -- and/or extract
all information on the lens potential from extended lensed images. The
general problem, however, is how to solve {\sl simultaneously} for
both the structure of the lensed source {\sl and} for the structure of
the lens potential (or the lens mass distribution). Information on
both are entangled in the complex structure of the multiple lensed
images.

\sn A significant step forward in disentangling this information and
using it to constrain the lens potential, is the use of non-parametric
image reconstruction techniques in which the source brightness
distribution is reconstructed on a, in some cases adaptive, grid
(e.g.\ Kochanek et al. 1989; Wallington, Kochanek \& Koo 1995;
Ellithorpe, Kochanek \& Hewitt 1996; Wallington, Kochanek \& Narayan
1996; Warren \& Dye 2003; Treu \& Koopmans 2004; Wucknitz 2004;
Wucknitz et al. 2004; Wayth et al. 2005; Dye \& Warren 2005; Brewer \&
Lewis 2005). These techniques fully exploit all information contained
in the lensed images. The use of smooth\footnote{i.e.\ slowly varying
on the scales of the lensed images.} parameterized potential models
ensures relative ease in the separation of information about the
source and the potential, since the degree of flexibility in the
potential models is significantly less than in the source model.

\sn To overcome a certain lack of freedom in the parameterized
potential/mass models, Saha \& Williams (1997) developed a non-parametric
mass reconstruction technique. Their implementation, however, only
uses constraints from point-like images and the number of free
parameters in the mass models often far exceeds the number of genuine
constraints. The interpretation of the results must therefore be done
statistically (e.g.\ Williams \& Saha 2000). Even so, the method is
powerful in exposing the range of mass models (although some might not
be physical) for systems with only lensed point images.

\sn {\sl Gravitationally lensed images, however, are strongly
correlated representations of the same lensed source, whereas the
effect of the lens potential on them should be highly un-correlated on
small scales. This allows them to be non-parametrically separated to a
very large degree.} 

\sn To illustrate this point more vividly, suppose that one observes a
two or four-image lens system.  Any change in the source structure
will show up in {\sl all} multiple images. Conversely, similar
structure in the lensed images most likely arises from structure in
the source and not from the unlikely coincidence of nearly-identical
(small-scale) perturbations in the lens potential that happen to occur
exactly in front of all lensed images. Contrary, clear structure in
only one, hence not all, of the lensed images is most easily explained
by a localized perturbation of the lens potential near that particular
image. By casting this in likelihood terms, this illustration tells us
that correlated structure in the lensed images -- even though not
excluded to arise from small-scale perturbations in the lens
potential in front of all lensed images -- is most likely the result
from structure in the source. Likewise, uncorrelated structure in the
lensed images is most likely the result of small-scale perturbations
in the lens potential.  This forms the basis of a maximum-likelihood
technique that allows the source and potential information encoded in
the multiple lensed images, to be separated. This does not imply,
however, that all degeneracies are broken (e.g.\ the mass-sheet
degeneracy; Falco, Gorenstein, \& Shapiro 1985). It does, however,
allow greater freedom in the lens potential models and the possibility
to reconstruct small scale potential perturbations, arising from
mass-substructure, if present.

\sn In this paper, we have developed this notion further and
extensively modify the non-parametric source reconstruction algorithm
-- as implemented in Treu \& Koopmans (2004)\footnote{We note that the
implementation of the semi-linear inversion technique in Warren \& Dye
(2003) is similar to that in Treu \& Koopmans (2004) and does {\sl
not} require the inversion of the lens equation as previously
suggested. Both implementations, however, differ in their
determination of the lens and blurring operators (Simon Dye, private
communications). Regularisation is done through a quadratic term added
to the $\chi^2$ penalty function, which ensures linearity of the set
of equations whose solution minimizes the penalty function
(maximum-likelihood or entropy terms are not quadratic and lead to a
non-linear set of equations that needs to be solved iteratively).} --
to allow for a simultaneous non-parametric source and lens potential
reconstruction. This integrates the idea initially suggested by
Blandford et al. (2001) and more recently by Suyu \& Blandford (2005),
with the semi-linear inversion technique by Warren \& Dye (2003) as
implemented in Treu \& Koopmans (2004) into a single mathematical
framework. Through the two-dimensional Poisson equation, the lens
potential provides a direct {\sl gravitational image} of the surface
density of the lens galaxy, not affected (or potentially biased) by
external shear.

\sn The technique of {\sl gravitational imaging} allows dark-matter
substructure in cosmologically distant lens galaxies to be detected,
imaged and quantified. Flux-ratio anomalies have already hinted at its
existence, either dark or luminous (e.g. Mao \& Schneider 1998;
Metcalf \& Madau 2001; Chiba 2002; Metcalf \& Zhao 2002; Dalal \&
Kochanek 2002; Brada\v{c} et al. 2002; Keeton 2003). If indeed these
anomalies are caused by perturbations of the underlying smooth
lens-galaxy potential, a direct image of the lens-galaxy mass
distribution could settle many of the still open questions, such as
which mass scale contributes most to flux-ratio anomalies and whether
this mass is associated with the lens galaxy (Brada\v{c} et al. 2004;
Mao et al. 2004; Amara et al. 2005; Metcalf 2005).

\sn Moreover, the unequivocal detection of dark-matter substructures
around lens galaxies -- without any reasonable mass-to-light ratio
counterpart in deep high-resolution optical images -- would be a
direct vindication of the basic cold-dark-matter paradigm that
predicts them to exist in large numbers (Moore et al. 1999; Klypin et
al. 1999). Because substructure is expected to be more prevalent at
higher redshifts (e.g.\ Gao et al. 2004), lens galaxies are excellent
probes and gravitational imaging an excellent way to test and quantify
the basic assumptions of the dark-matter and the hierarchical
structure formation models.

\sn In Section 2, we derive an iterative set of linear equations that
solve {\sl simultaneously} for the source structure and the lens
potential. We also discuss which form of regularization to use and how
to determine the regularization parameters through a Maximum
Likelihood analysis. In Section 3, we illustrate the method with a
simulated Einstein ring. In Section 4, we draw conclusions.

\begin{figure*}
\begin{center}
\leavevmode 
\hbox{%
\epsfxsize=0.90\hsize 
\epsffile{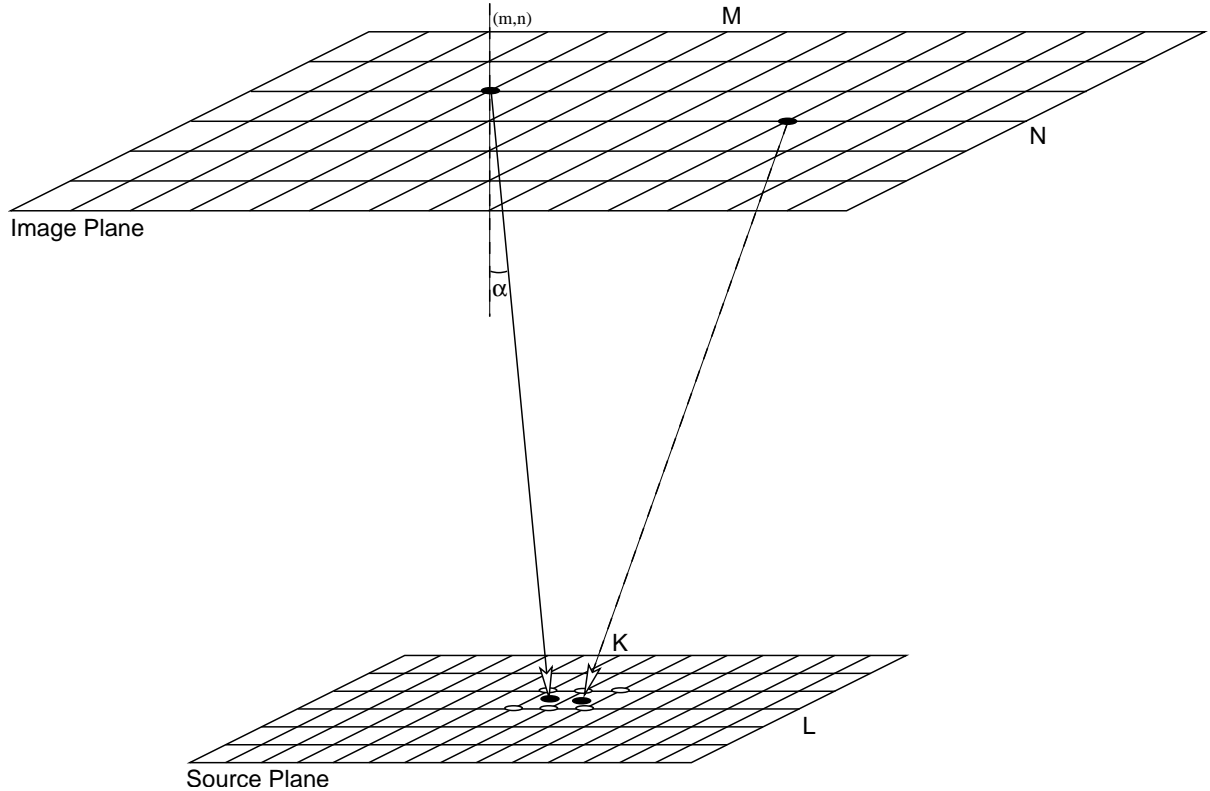}}
\end{center}
\caption{A schematic overview of the method of non-linear image (i.e.\
source) reconstruction, as implemented by Treu \& Koopmans (2004) and
in this paper. In the absence of blurring or averaging inside pixels,
the position ($\vec x$) of each pixel $(m,n)$ in the image plane
corresponds to a position in the source plane ($\vec y$), through the
lens equation $\vec y = \vec x -\vec \alpha(\vec x)$. The surface
brightnesses at these corresponding points -- conserved through
lensing -- are the same. Because the source brightness distribution is
reconstructed on a fixed grid, the surface brightness at $\vec y$,
i.e.\ $\Sigma(\vec y)=\Sigma(\vec x)$, is represented by a linear
superposition of the surface brightnesses at the four pixels that
enclose $\vec y$ (open circles). The weights for each of these source
pixels are the bilinear interpolation weights, whose sum add to unity
to conserve flux (higher-order interpolation is also possible; see
Treu \& Koopmans 2004). This way -- because of the multiple nature of
the lensed images -- there can be more than one constraint on a
single source pixel (depending on its size and the number of multiple
images). In addition, because there are multiple solutions of four
weighted brightnesses adding to the observed brightness at $\vec x$,
regularisation is often required to ensure a relatively smooth and
more physical source brightness distribution for lower signal-to-noise
data (see text).}
\label{fig:method}
\end{figure*}

\section{The Method}

In this section, before we extend the non-parametric method to include
a non-parametric reconstruction of the lens potential, we first
discuss the non-parametric reconstruction method for only a pixelized
source, but given a smooth parameterized lens-potential model.

\subsection{The non-parametric source solution}\label{sect:sourcesol}

The non-parametric reconstruction method of a pixelized source is
shown schematically in Fig.~1. The figure illustrates how the surface
brightness of each image pixel can be represented through a weighted
{\sl linear} superposition of the (unknown) surface brightnesses at
four source pixels. Hence, one can represent this as a simple linear
equation (see Koopmans \& Treu 2004) and because this holds for each
of the M$\times$N image pixels, one obtains a set of M$\times$N
coupled\footnote{The brightnesses of multiple image pixels can depend
on the brightness of a single source pixel (see the caption of
Fig.~1).} linear equations. This set of equations is constrained by the
M$\times$N observed surface brightnesses values (i.e.\ $\data$) and
has K$\times$L free parameters (i.e.\ $\source$; the unknown surface
brightnesses values on the source grid).

\sn As was shown in Koopmans \& Treu (2004) and also Warren \& Dye
(2003), the pixelized lensed image (with the lens-galaxy subtracted
and blurring included) can then be expressed as the set of linear
equations
\begin{equation}\label{eq:lineq_source}
  \BO \LO(\psi)\, \source = \data,
\end{equation}
where $\LO(\psi)$ and $\BO$ are the lensing and (also linear)
blurring operators, respectively, and $\psi$ is the lens potential
which for simplicity is here also representing the set of unknown
lens-potential parameters. Each row of the lensing operator (a sparse
matrix) only contains the four bi-linear interpolation weights, placed
at the columns that correspond to the four source pixels that enclose
the associated source position (Fig.~1; see Treu \& Koopmans 2004 for
details).

\sn The set of equations (1) is in general ill-posed (i.e.\
under-constrained and/or noisy data) and can not be solved through
$\source = [\BO \LO(\psi)]^{-1}\data$. Instead, there exist well-known
regularized inversion techniques that minimize the quadratic penalty
function
\begin{equation}
  G(\source, \psi) = || \OO  \source - \data||_2^2 + \sum_i
  \lambda_i || {\bf H}_i \source||^2_2,
\end{equation}
by varying $\source$ (we define $\OO\equiv\BO\LO (\psi)$). In
addition, ${\bf H}_i$ are regularization matrices and $\lambda_i$ are
the corresponding regularization parameters, both of which we discuss
in more detail below (see also Press et al. 1992 for a clear
discussion about regularization methods). The first term in the
penalty function $G(\source, \psi)$ is proportional to the total
$\chi^2$ for a given solution $\source$ and as such quantifies how
well the model matches the data. The second terms can regularize the
``smoothness'' of the solution by adding a positive quadratic penalty
term, whose argument depends linearly on the source solution, i.e.\
through ${\bf H}_i \source$, which for example could give its $n$-th
order derivative. The quadratic nature of the regularization term
ensures that the minimum--$G$ solution can be found by solving a
linear set of equations (this is not the case for maximum-entropy or
maximum-likelihood regularization terms). The quadratic nature of the
regularization term dramatically reduces the time to numerically find
the best source solution and ensures that the global minimum of $G$ is
found (given a fixed $\psi$).

\sn By setting the derivative $\partial G(\source, \psi)/\partial \vec
s =0$, a bit of linear algebra then leads to the required set of equations
\begin{equation}\label{eq:newset}
  \left[ \OOT \OO + \sum_i \lambda_i {\bf H}^{\rm T}_i {\bf
  H}_i\right] \source = \OO^{\rm T} \data,
\end{equation}
whose solution for $\source$ minimizes $G(\source,\psi)$, by
construction, for a fixed lens potential $\psi$. In an outer
non-linear optimization loop (see Warren \& Dye 2003) one then varies
the free parameters of $\psi$ to find the {\sl global} minimum of the
joint penalty function, $G(\source, \psi)$, resulting in an optimized
non-parametric source model and an optimized set of parameters 
for the lens potential.

\sn In this way, one makes full use of all information in the
lensed images to constrain the parameterized potential model $\psi$,
without strong assumptions about the structure of the source. However,
there are several questions that arise:\smallskip

\smallskip\noindent {\bf (1)} Does the parameterized potential model
$\psi$ have enough freedom to describe the true lens potential? 

\smallskip\noindent {\bf (2)} Which form for the regularization
matrices, ${\bf H}_i$, should be used?

\smallskip\noindent {\bf (3)} What should be the values of the
regularization parameters $\lambda_i$ in the penalty function $G$?

\smallskip\noindent To (partly) address these questions, we derive in
section \ref{sect:lincorr} a {\sl new} set of linear equations, whose
min--$G$ solution gives the pixelized source structure and
simultaneously a linear pixelized correction ($\delta \vec \psi$) to
the initial model of the lens potential. By solving this new set of
equations and correcting the potential model iteratively, one
minimizes the penalty function to find a {\sl non-parametric and
non-linear} solution for both the source structure {\sl and} the lens
potential.

\sn We choose to pixelize and solve for the lens potential, in
contrast to e.g.\ Saha \& William (2000) who solve for a pixelized
mass distribution. Both are related through the two-dimensional
Poisson equation. However, because {\sl all} properties of the lens
system can be derived solely from the {\sl local} lens potential
(through its zero-th, first and second order derivatives), {\sl no}
assumptions have to made about what happens outside the region where
we have lensing information (i.e.\ the lensed images). On the other
hand, if one chooses to use a pixelized mass distribution, one needs
to make some a priori assumptions about the mass distribution outside
the grid (e.g.\ no mass), since mass {\sl outside} the grid still
contributes to the lens potential {\sl inside} the grid. If wrong
assumptions are made, for example one assumes there is no potential
arising from external shear (i.e.\ $\psi_{\rm sh}(\vec x)=0$) whereas
in fact there is, one can severely bias the resulting solution for the
pixelized mass distribution, because the contribution from shear must
then somehow be mimicked by the mass distribution inside the grid. 

\sn In the case of a pixelized potential model, however, the lens
potential due to external mass (i.e.\ from outside the grid) would be
reconstructed without a problem, just like that arising from mass
inside the grid. The Poisson equation, furthermore, only gives the mass
distribution inside the grid, not affected by e.g.\ external shear
(because $\nabla^2 \psi_{\rm sh}(\vec x) = 0$) or any mass external to
the grid. The reconstructed mass distribution can therefore never be
biased by assumptions about the mass distribution outside the
grid. We regard the latter as a great advantage over the use of a
pixelized mass distribution.

\sn A second advantage of using the lens potential over the lens mass
distribution is increased numerical speed in the optimization
process. In case a mass distribution is used, the lens potential still
needs to be calculated through a convolution. This is a significant
numerical burden, whereas all lens properties can simply and quickly be
derived from a lens potential through its $n$--th order derivatives
($n$=0, 1 or 2).

\sn Despite these advantages, the use of a pixelized mass distribution
provides a means of retaining a positive surface density. However,
when we regularise the potential model by minimizing the norm of its
$n$-th order derivative, we regularize the mass density in its
($n$$-$2)--th order derivative (because of the second order derivative
in the Poisson equation). Because we start with a smooth
positive-density mass model and only make a relative small potential
correction, the regularization can thus ensure that the surface
density of the mass model remains positive, even though we reconstruct
the lens potential (see also~\S \ref{sect:regul2}).

\sn The second and third questions, stated above, are of more general
interest in many similar (image-reconstruction) problems and are
addressed in sections \ref{sect:regul} and \ref{sect:GML}, within the
context of the problem at hand.

\begin{figure*}\label{fig:true}
\begin{center}
\leavevmode \hbox{%
\epsfxsize=0.90\hsize 
\epsffile{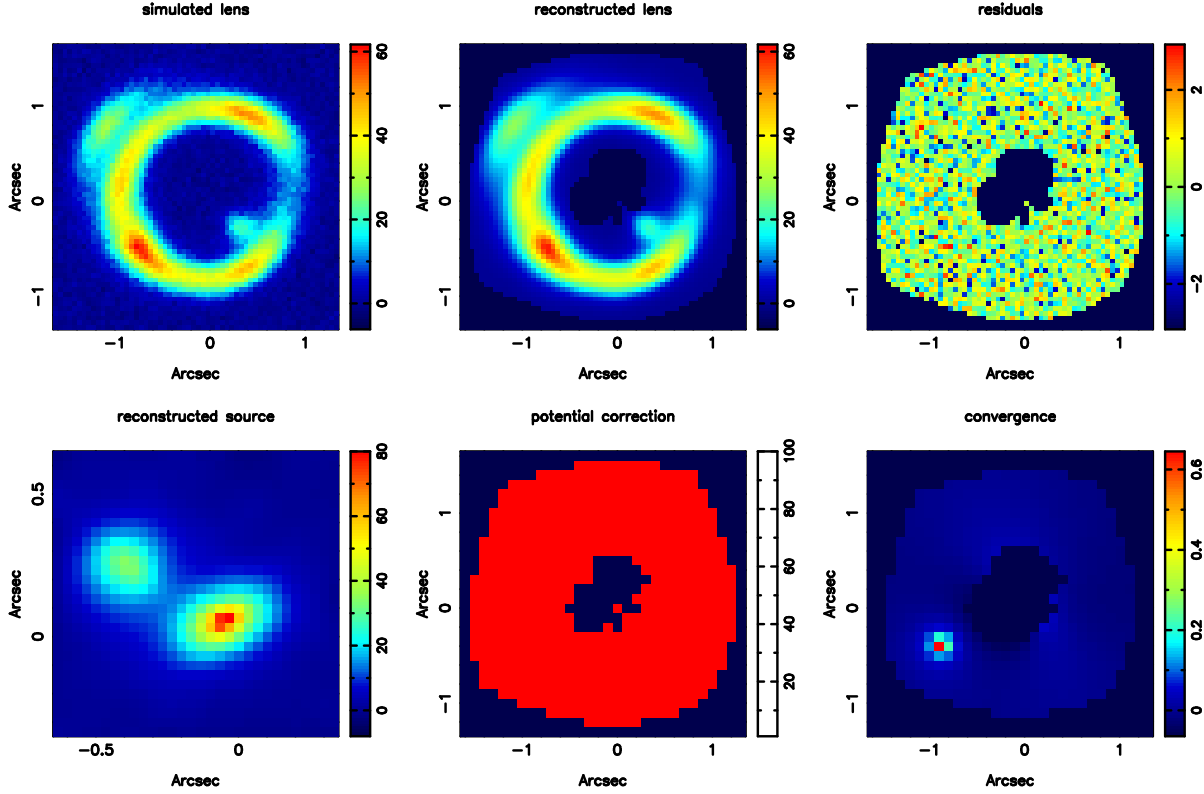}}
\end{center}
\caption{Reconstruction of the source and lens-potential correction
using the true mass model. {\bf Top row:} A simulated Einstein-Ring
lens system (see text) (left). The best reconstruction (middle) and
residuals (right). {\bf Bottom row:} The reconstructed source
(left). The correction to the lens potential, set to zero for this
model (middle), and the total convergence minus a best-fit SIE mass
model (right). The shape of the reconstructed region is determined by
the projection of the rectangular source plane on to the image
plane. This results in a hole in the center of the reconstruction,
since the source plane (in this case) maps on to a closed
annulus. Outside the annulus, image-plane pixels map outside the
source-plan grid.}\label{fig:true}
\end{figure*}

\subsection{The non-parametric source and potential 
solutions}\label{sect:lincorr}

To find a linear correction to, say, a parameterized potential model
(e.g.\ found as described in \S\,\ref{sect:sourcesol}) we assume that
the potential $\psi$ is also pixelized\footnote{Because we determine a
correction to the initial potential model -- which itself might be
parametrized -- we in principle only need to pixelize the potential
correction and not the initial $\psi$. The result, however, is
the same.}  (i.e.~$\pot$ are the pixel values). We note that the
number and size of the pixels for the potential grid need not be the
same as for the image grid.

\sn Given the previously-found best-fit source model $\source_{\rm p}$
and potential-model $\pot_{\rm p}$, one can then subtract the best-fit lens
model from the data and obtain a residual image
\begin{equation}\label{eq:resid1}
  \delta \vec{d} = \data - \OO(\pot_{\rm p}) \, \source_{\rm p}.
\end{equation}
If $||\delta \vec{d}/\sigma_{d}||^2\gg (MN)$, where $\sigma_d^2$ is
variance per image pixel, either the source or the potential
model is not correct, or both. We can then assume that a potential
correction, $\dpot$, exists for which to first order
\begin{equation}\label{eq:resid2}
  \data - \OO(\pot_{\rm p}+\dpot) \, \source_{\rm p} \rightarrow \vec{0}.
\end{equation}
It is our task to find $\dpot$ and with it correct the previously
best-fit potential model $\pot_{\rm p} + \dpot  \rightarrow \pot$. With the
previous source solution, $\source_{\rm p}$, this can to first
order  also be written as new set of linear equations (Appendix A)
\begin{equation}
  \delta \vec{d} = - \BO \DS \DP \dpot,
\end{equation}
where $\DS$ is a sparse matrix whose entries depend on the gradient of
the previously-best source model and $\DP$ is a matrix that determines
the gradient of $\dpot$.  Note also that the image blurring is
accounted for (see Koopmans \& Treu 2004). Combining equation (6) with
equation (4), one finds
\begin{equation} \label{eq:firstorder}
  \BO \left[ \LO(\pot_{\rm p}) \source - \DS \DP \dpot\right] = \data
\end{equation}
If we further introduce the block matrix 
\begin{equation}
  \CO \equiv \left( \begin{array}{ccc}
       \LO(\pot_{\rm p})\mid - \DS \DP 
       \end{array}
  \right)
\end{equation}
and the block vector
\begin{equation}
  \vd \equiv \left( \begin{array}{c}
    \source\\
    \dpot
    \end{array}
  \right),
\end{equation}
the final linear set of equations becomes
\begin{equation}\label{eq:srcpot}
 \BO\CO\,\vd \equiv \MO\, \vd = \data
\end{equation}
The structure of this equation is the same as that of
equation~\ref{eq:lineq_source} and can thus be solved as before. We
note that because the first term of equation~\ref{eq:firstorder}
depends only on $\pot_{\rm p}$ and $\vec s$, whereas the second terms
depends only on $\dpot$ and $\vec{s}_{\rm p}$, they are decoupled sets
of equations that could be solved independently.

\sn How can the correction in equation (6) be interpreted?  This is
most easily seen when writing the equation in its continuous form
$\delta d(\vec{x}) = - \vec \nabla_y s(\vec y) \cdot \vec \nabla_x
\delta \psi(\vec{x})$ as shown in Appendix A. Because $\vec y = \vec x
- \vec{\nabla} \psi$ and $\delta \vec y = - \vec{\nabla} \delta \psi$,
what equation (6) tells us is that to first order the deviation of
$\delta d(\vec x)$ from zero can be compensated exactly by Taylor
expanding $s(\vec y)$ to first order around $\vec y$ and then changing
the deflection angle slightly by $\delta \vec y$ to point to that part
of the source, close to $\vec y$, that has the same surface brightness
as that seen at the image position $\vec x$.  The trick is to find the
solution $\delta \psi(\vec{x})$ such that $\delta d(\vec{x})
\rightarrow 0$ for all $\vec x$. This is what has been done above, by
linearizing the equations and writing the result in matrix/vector notation.

\begin{figure*}
\begin{center}
\leavevmode 
\hbox{%
\epsfxsize=0.45\hsize 
\epsffile{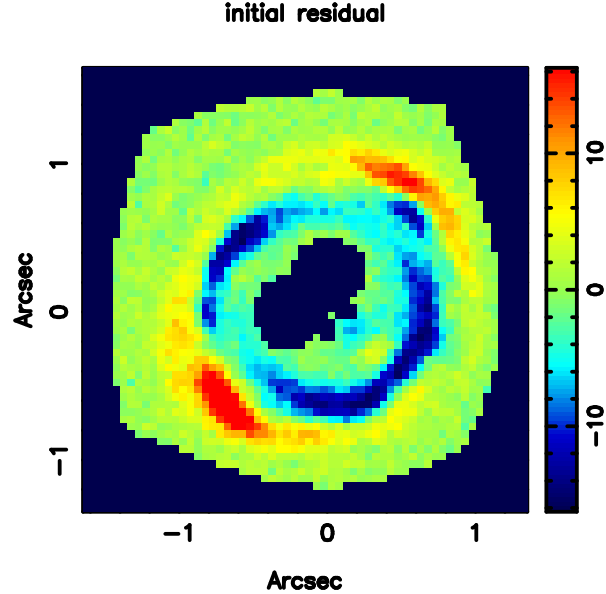}}
\end{center}
\caption{The residuals of the data minus the ``best'' reconstruction,
assuming a SIE mass model fit to the image centroids only. Note the
difference with the final residuals in Figure~\ref{fig:sim1}.}
\label{fig:resid}
\end{figure*}

\subsection{The Regularization Matrix}\label{sect:regul}

Because $\source$ and $\dpot$ are independent, they require their own
regularization parameters. Consequently, the regularization matrix needs
to be modified somewhat:
\begin{equation}
  {\bf R} \equiv \left(
  \begin{array}{cc}
    \sum_j \lambda_{s,j} {\bf H}^{\rm T}_{s,j} {\bf H}_{s,j} & 0 \\ 0
    & \sum_j \lambda_{\delta \psi,j} {\bf H}^{\rm T}_{\delta \psi,j}
    {\bf H}_{\delta \psi,j}
  \end{array}
  \right).
\end{equation} 
The set of equations, whose solution minimizes the penalty function,
then becomes
\begin{equation}\label{eq:srcpotreg}
\left( \MOT \MO + {\bf R}\right)
  \vd = \MOT \data.
\end{equation}

\sn Hence, by iteratively solving equation~(\ref{eq:srcpotreg}) and
correcting the lens potential at each iteration, one finds the source
and potential structure that minimize $G$ without strong assumptions
about either. This equation is the equivalent of equation (3).

\subsubsection{Regularizing with higher-order derivative operators}
\label{sect:regul2}

\sn The form of the regularization matrices is not a-priori
determined. But since ``smoothness'' is often a criterion for how well
the reconstruction has been done, most often derivative operators are
used for the regularization matrices (see Press et al. 1992 for some
examples); the zeroth order derivative being the identity
matrix. Their mathematical structure is also the best understood of all
forms (see Neumaier 1998 for a complete analysis).

\sn We use different derivative-orders for the source and the
potential regularization. We find that zeroth order derivative, ${\bf
H}={\bf I}$, gives poor source reconstruction in the majority of
cases, especially if the signal-to-noise ratio of the data is low (see
Treu \& Koopmans 2004 for some discussion) and worse for the potential
reconstruction, that requires smoothness in very high order (see
below). One solution is to resort to {\sl multi-scale} pixels in the
source plane (Dye \& Warren 2005), which is an implicit form of
adaptive regularization. But its hard to require smoothness (or
continuity) in the solutions and their derivatives.

\sn For this reason, a much better choice of regularization for
{\sl single-scale} pixels, is the use of higher-order derivative operators.
In those cases neighboring pixels are ``connected'', resulting in
considerably improved solutions\footnote{Of course, {\sl any} choice of
regularization is {\sl ad hoc} at some level. Only by testing the
different choices can one assess whether the regularization gives a
(physically) acceptable solution.}.

\sn It is found that regularization with a single second-order derivative
operator (i.e.\ minimizing curvature) is often sufficient to
reconstruct the source. To require the convergence (i.e.\ surface
density) of the lens to be smooth, the potential regularization
requires {\sl at least} a single fourth-order derivative
operator\footnote{The convergence is derived from the potential using
second-order derivatives in the Poisson equation.}. Since $\delta
\kappa/\kappa \ll 1$ and $\delta\kappa$ is smooth for any good
starting model, one ensures in general positive convergence solutions.

\begin{figure*}
\begin{center}
\leavevmode
\hbox{%
\epsfxsize=0.90\hsize 
\epsffile{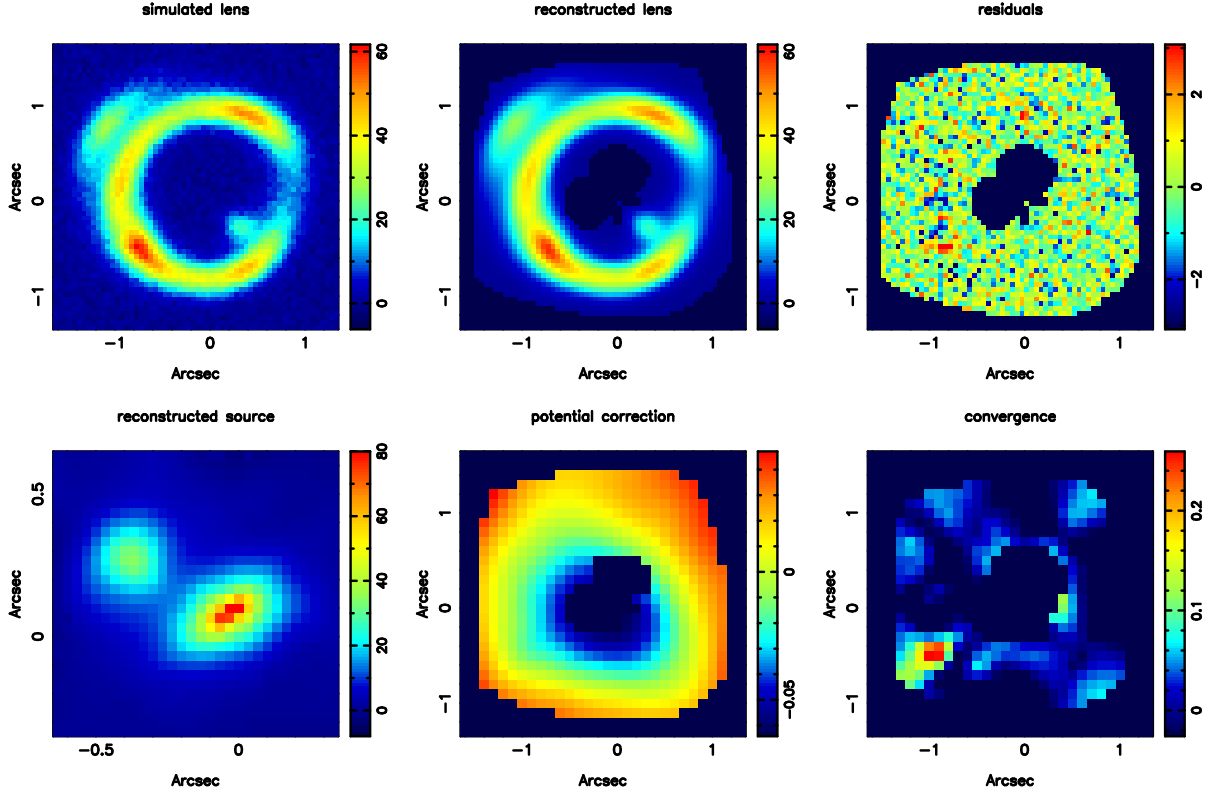}}
\end{center}
\caption{Reconstruction of the source and lens-potential correction
using a best-fit single SIE mass model as starting model. {\bf Top
row:} A simulated Einstein-Ring lens system (see text) (left). The
best reconstruction (middle) and residual images (right). {\bf Bottom
row:} The reconstructed source (left). The correction to the lens
potential (middle) and the lens convergence minus a best-fit SIE mass
model~(right).}\label{fig:sim1}
\end{figure*}

\subsection{The Regularization-Parameter Values}\label{sect:GML}

\smallskip \noindent Having chosen appropriate regularization
matrices, the next question is which level of regularization is
appropriate.  

\sn {\bf (i)} The ``subjective'' way of choosing their numerical
values is through a careful analysis of simulated lens systems that
closely resemble the real lens system. The regularization-parameters
values are then chosen such that the input models are best
recovered. Because the real system resembles the simulated system, one
assumes that the obtained value for the regularization parameter also
gives an unbiased reconstruction of the source and potential of the
real system.

\sn {\bf (ii)} The ``objective'' way is to have the data itself
determine the regularization-parameter values. Intuitively, one might
think that the probability distribution of pixel values of the
difference between the observed system and the model should be that of
``noise'' with zero mean and show no residual structures. How can in
that case, the numerical values of $\lambda_{s/\delta \psi, j}$ be
determined?

\sn As often is the case in CCD images, if the noise is approximately
Gaussian distributed, this can be expressed for a fixed $\psi$, as the
maximization of the likelihood
\begin{equation}\label{eq:lh1}
  {\cal L} = \frac{1}{(2 \pi)^{NM/2} \sqrt{{\rm det}(\sigma_{d}^2 {\rm {\bf
  I}})}} \exp\left[-(\data - \OO \tilde {s})^{\rm T} (\data - \OO
  \tilde {s}) /(2\, {\sigma_d^2})\right],
\end{equation} 
with $\sigma_d^2$ being the variance of the image pixels, assuming
that the covariance between different pixels is zero, and $\tilde {s}$
being the solution from equation (3). What this equation means is that
the maximum-likelihood value of $\lambda$ should, as closely as
possible, result in a zero-mean Gaussian residual image $\delta
\data$. As an illustration, if $\lambda$ (encoded in $\tilde s$) is
too small, the model will fit the noise, resulting in a ``flat''
residual distribution function (i.e.\ the lowest noise values becomes
part of the model and consequently don't show up in the residual
image). On the other hand, if $\lambda$ is too large, the residual
image will show coherent non-Gaussian structures that could not be
fitted because of the over-regularization. In both cases, the
residuals deviate from Gaussian, resulting in a low likelihood for the
model.

\sn Using the estimate of $\tilde s$ from equation (3), it has been shown
shown by Wahba (1985; see also Neumaier 1998) that the Maximum
Likelihood value of the regularization parameter for the source can
then be found by maximizing the likelihood function,
\begin{equation}\label{eq:lh2}
   {\cal L} =  \log(||\data||_2^2 - ||\vec{u}||_2^2) - \log(\lambda) 
  +\frac{2}{n} \log {\rm det}({\bf C}),
\end{equation}
where ${\bf C} {\bf C}^T =\OOT \OO + \lambda\, {\bf H}^{\rm T} {\bf
H}$ is obtained through a Cholesky decomposition, $\vec{u}$ is found
by solving ${\bf C} \vec{u} = \OOT \data$ and $n=KL$ is the number of
source pixels. The result is independent from the variance
${\sigma_d^2}$, which therefore does not have to be explicitly known a
priori. The ML solution of the source is obtained by solving ${\bf
C}^T \source = \vec{u}$. Hence equation \ref{eq:lh2} is the equivalent
of equation \ref{eq:lh1}, but now explicitly expressed with $\lambda$
and in such a way that it is easy to solve numerically through a
Cholesky decomposition that also gives the solution for $\source$
without much additional effort (i.e.\ through a fast back-substitution,
because ${\bf C}$ is a diagonal matrix).

\smallskip \noindent Since we solve iteratively for the lens
potential, it is found that the value of the regularization for the
potential is almost irrelevant if chosen large enough (it just takes
longer to converge for stronger regularization). Furthermore, since
the solutions of $\vec{s}$ and $\pot$ are {\sl decoupled} in the set
of linear equations, as we discussed above, determining the value
of $\lambda$ for the source separately from the potential is mathematically
correct. 


\section{Gravitational Imaging of Galaxy (Sub)structure}\label{sims}

Having introduced the methodology above, we refer to appendix~B for an
outline an iterative algorithm of the method. Although it is not
necessarily the only or most optimal implementation, it is relatively
easy to implement and robust. To test the algorithm, we performed a
simulation of an artificial Einstein-ring and subsequently reconstruct
the source and lens potential.

\sn In \S 3.1, we first describe the simulated lens system. The source
is chosen to be a double structure, consisting of two elliptical
galaxies.  Similarly, the lens is a smooth elliptical mass
distribution plus an additional low-mass substructure. The aim of this
simulation is to show that the method/algorithm is capable of
reconstructing both the source structure and the small-scale mass
distribution (i.e.\ the mass substructure) on the lens galaxy.

\subsection{Artificial Einstein-ring Lens System}

\sn The components of the lens-system are:

\sn {\bf (1) Lens Mass Model:} The lens mass model consists of a
single SIE lens (Kormann et al. 1994) with a lens strength of
$b$=$0\farcs9$, a position angle (PA) of $\theta$=$45^\circ$, an axial
ratio of $f$=$(b/a)_{\kappa}$=$0.8$ and centered at $(0\farcs0,
0\farcs0)$. The mass model also includes a dwarf-satellite represented
by a SIS with $b=0\farcs045$ at $(-0\farcs9, -0\farcs4)$ (placed on
the Einstein ring). Since typical lens galaxies with these image
separations have $M_{\rm E}$$\sim$$10^{11}$~M$_\odot$ inside their
Einstein radius, this particular substructure has a mass of $M_{\rm
E}$$\sim$$10^8$~M$_\odot$.

\sn {\bf (2) Source Brightness Model:} To show that complexity
in the source and lens-potential models can be disentangled, the
source consist of two components: (1) A subcomponent with an
elliptical exponential brightness profile with $0.1$--arcsec
scale-length, a central surface brightness of 100~arcsec$^{-2}$
(arbitrary units), an axial ratio of 0.64, a PA=113$^\circ$ and
centered at $(-0\farcs05,0\farcs05)$.  (2) A subcomponent with an
exponential brightness profile of $0.1$--arcsec scale-length, a
central surface brightness of 50\,arcsec$^{-2}$, centered at
$(-0\farcs40,0\farcs25)$. The source is pixelized on a
1\farcs0$\times$1\farcs0 grid of $30\times 30$ pixels.

\sn {{\bf (3) Simulated Lensed Images:} The lensed image is
calculated on a 3$''$$\times$3$''$ grid of 60$\times$60 pixels and
blurred by an artificial HST--ACS F814W PSF. Gaussian noise with
$\sigma=1$ is added to the resulting blurred image. The simulated
system and reconstructed source model are shown Fig.\ref{fig:true}.

\subsection{The Non-Parametric Source and Lens-Potential Reconstruction}

\sn We perform two reconstructions. The first is a test-run and
reconstructs {\sl only} the source structure, assuming that we have
perfect knowledge of the true lens mass model.  This is not a
realistic situation, but it shows that the method properly recovers the
input source model without significant residuals. The potential grid
is defined on a 3$''$$\times$3$''$ grid of 30$\times$30 pixels,
sufficient to capture (sub)structure in the lens potential. The result
is shown in Figure~1. The input source and Einstein ring are nicely
reconstructed and the residuals are not significant ($\chi^2/{\rm
NDF}$=0.96). 

\sn The second run should represent a more realistic situation. As
would be done with an observed lens system, we find the source and
lens-potential model in a number of distinct steps: (1) First, we fit
a single SIE mass model to the four image centroids; the resulting
best-fit parameters are $b$=$0\farcs85$, a PA of $\theta$=$47^\circ$,
an axial ratio of $f$=$(b/a)_{\kappa}$=$0.84$). The residuals have
$\chi^2$/NDF=29.4, hence the model is dramatically far from the true
mass model (see Figure~\ref{fig:resid}). Although statistically a very
poor model, we can use this as the initial starting model for the
non-parametric reconstruction. (2) We non-parametrically reconstruct
the source and the lens-potential correction, with a regularization
parameter for the source ($\lambda_{\rm s}$=3.0; which is clearly too
large) and the potential ($\lambda_{\delta \psi}=10^9$),
respectively. We lower the regularization parameters of the potential
by 0.1 of its previous value each iteration and iterate $\sim$60
times, until convergence ($\sim$30 minutes on a 3--GHz laptop). (3)
Using this solution, which is an extreme improvement
($\chi^2$/NDF=1.19), we use the ML technique (Section \ref{sect:GML})
to determine the value of the source regularization parameter,
$\lambda^{\rm ML}_{\rm s}=0.4$. (4) We rerun the simulation with the
new value of the regularization parameter. We note that the precise
value of $\lambda_{\rm s}$ in the second run is nearly independent
from its value in the first run, justifying this approach.

\sn The best model has $\chi^2$/NDF=1.05 and is shown in
Fig.~\ref{fig:sim1}, a dramatic improvement over the best-fit single
SIE mass model.  We subtract a best-fit single SIE mass model (fitted
to the total convergence; see lower-right panel) to highlight any
substructure of the lens. The small mass-perturbation is indeed
recovered near the correct position and with good dynamic range,
compared with the residuals in the convergence field. We note that
regularization smooths the structure of the small mass component,
which is unavoidable. However, within a $0\farcs7\times 0\farcs7$
aperture centered on the substructure (approximately the region of
positive convergence after the smooth model was subtracted), we
recover the original mass of the substructure to within 12\%. This
shows that one can not only detect, but also quantify its mass.

\sn We note that the final solution has positive convergence
everywhere inside the grid. In Figures 2 and 4 only the convergence is
show after subtracting a smooth SIE model, hence the residual from
this could be negative (as opposed to the total convergence).

\sn Although we only presented a single simulation in this paper, we find
that the algorithm can also reconstruct both the source and the lens
potential in more complex and lower S/N cases. A more thorough
analysis of the method, its errors and degeneracies, however, is
planned. It should include blind tests on N-body simulated lens
systems and the use more realistic source structures.

\section{Conclusions}

A new non-parametric source {\sl and} lens-potential reconstruction
method has been presented and implemented. The method has been used to
{\sl gravitational image} substructure in an artificial (but
reasonably realistic) lens system, recovering the position and
enclosed mass of the substructure, as well as the structure of the
lensed source. We also conclude that the method can disentangle
structure in the source and lens potential, without strong assumptions
about either.

\sn Further improvements, however, are still required: (i)
Determination of the full covariance matrix to determine the errors
and correlations between the source brightness pixels and
potential-correction pixels. (ii) The use of a Markov--Chain
Monte--Carlo technique to determine the correct (non-linear) errors
for each of the source and potential pixel values. This provides a way
of estimating errors in real lens systems where Monte-Carlo
simulations are not feasible. (iii) Testing multi-scale pixel schemes
(Dye \& Warren 2005) against single-scale pixel schemes with adaptive
regularization. Due to severe computational limits, however, we have
not yet implemented these.

\sn Having demonstrated its feasibility, {\sl gravitational imaging}
can serve as a new tool to discovery {\sl and} quantify the level and
evolution of the (dark-matter) substructure in the halos around
galaxies at cosmological distances. Through it, the cold-dark-matter
paradigm and the hierarchical structure formation models can be
tested.

\section*{Acknowledgments} The author would like to thank Roger Blandford,  
Maru\v{s}a Brada\v{c}, Simon Dye, Chris Fassnacht, Phil Marshall,
Sherry Suyu, Peter Schneider, Tommaso Treu and Saleem Zaroubi for
useful discussions. The author also thanks the referee, Steve Warren,
for very helpful comments and suggestions that further improved the
presentation and clarity of the paper.

\appendix

\section{Linear correction of the lens potential}
Given the best-fit solution to equation \ref{eq:newset} for both the
potential $\psi$ and source $\source$, one can
determine the residual vector
\begin{equation}\label{eq:resid3}
  \delta \vec{d} = \data - \OO(\psi) \, \source.
\end{equation}
Hence one subtracts the best lensed source model (blurred with the
PSF) from the data. Suppose further that any significant residuals are
due to deviations of $\psi$ from the true potential (this can't
strictly be true, because $\source$ must also be slightly wrong if
$\psi$ is slightly wrong). The question arises which correction to
the potential, $\delta \psi$, we have to apply such that, {\sl after
the correction}, the residual vector becomes 
\begin{equation}
  \delta \vec{d} = \data -
  \OO(\psi + \delta \psi) \, \source \approx \vec{0}.
\end{equation}

\smallskip \noindent This can be derived as follows, for now regarding
the image, the source and the potential as scalar functions
$d(\vec{x})$, $s(\vec{x})$ and $\psi(\vec{x})$, respectively, and
neglecting the PSF. From a simple linear expansion of the lens
equation $\vec y = \vec x - \vec \nabla_x \psi(\vec{x})$
(e.g. Schneider et al. 1992) it then follows that
\begin{equation}
  \vec{y} +\delta \vec{y} = \vec{x} - \vec \nabla_x (\psi(\vec{x}) +
  \delta \psi(\vec{x}))
\end{equation}
or simply (for a fixed position $\vec x$)
\begin{equation}
  \delta \vec{y}  = - \vec \nabla_x \delta \psi(\vec{x}).
\end{equation}
Hence a change in the source position is related to a change in the
gradient of the lens potential. By adding a correction, $\delta
\psi(\vec{x})$, to the lens potential we can ``move around'' rays of
light hitting the source plane. Thus, our task is to find that
correction to the potential $\delta \psi(\vec{x})$, that obeys the
following relation: $s(\vec y+ \delta \vec y)- s(\vec y) = \delta d(\vec x)$
[i.e. the brightness of the lensed image at $\vec x$ equals the
brightness of the source at position $\vec x- \vec \nabla_x
(\psi(\vec{x}) + \delta \psi(\vec{x}))$].  The residual field is then
to first order
\begin{equation}
  \delta d(\vec{x}) = s(\vec{y} +\delta \vec{y}) - s(\vec{y}) \approx
  \vec \nabla_y s(\vec y) \cdot \delta \vec y.
\end{equation}
Combining equations (A4) and (A5), one finds
\begin{equation}\label{eq:pcorr}
  \delta d(\vec{x}) \approx - \vec \nabla_y s(\vec y) \cdot \vec
  \nabla_x \delta \psi(\vec{x}),
\end{equation}
to first order. This has previously been derived by Blandford
et al. (2001). The next step is now to derive the equivalent
equation for $\delta \vec{d}$ in equation \ref{eq:resid1}. 

\smallskip\noindent First, we assume that the potential grid has
$P\times Q$ pixels ($P$ columns and $Q$ rows) with $p=1\dots P$ and
$q=1\dots Q$. In addition, the potential values are $\psi_h$ with $h =
p + (q-1)P$ and $h=1\dots PQ$. The source and data-grids are defined
and described in Treu \& Koopmans (2004).

\smallskip\noindent We can now derive equation \ref{eq:resid1}. First,
we define
\begin{equation}
  \DS= \left( \begin{array}{cccccc} \dots & \\ & \frac{\partial
    s_h}{\partial y_1} & \frac{\partial s_h}{\partial y_2} \\ & &
    &\frac{\partial s_{h+1}}{\partial y_1} & \frac{\partial
    s_{h+1}}{\partial y_2} \\ & & & & & \dots\\
    \end{array}
  \right),
\end{equation}
where the entries indicate the gradient of source brightness
distribution in the $y_1$ and $y_2$ directions in the source
plane. The entries are evaluated at the potential-grid positions $\vec
y_h = \vec x_h -\vec \nabla \psi(\vec x_h)$ (the potential grid can be
different from the data grid).  In addition, if
\begin{equation}
  \DP \dpot= \left( \begin{array}{cccccc}
    \dots & \\
     & \frac{\partial \delta \psi_h}{\partial x_1} \\
     &  \frac{\partial \delta \psi_h}{\partial x_2} \\
     & & \frac{\partial \delta \psi_{h+1}}{\partial x_1} \\
     & & \frac{\partial \delta \psi_{h+1}}{\partial x_2} \\
      & & & & & \dots\\
    \end{array} 
  \right),
\end{equation}
It is then easy to see that $\delta \vec{d} = - \DS \DP \dpot$ (not
blurred) gives a vector whose entries are given by Eq.~\ref{eq:pcorr}.
The PSF smearing is simply included by multiplying with the blurring
operator $\BO$. The gradients can be evaluted through
finite-differencing schemes. Higher-order schemes will ensure
continuity in the derivatives of the source brightness distribution
and potentials.

\section{The Algorithm and Implementation}

By solving equation \ref{eq:srcpotreg}, adding the correction $\dpot$ to
the previously best potential model and iterating this procedure, both
the potential model $\pot$ and source model $\source$ should converge
to a solution that minimizes the penalty function~$G(\source, \psi)$. We find
the following algorithm to work very well, typically converging in $<$
100 iterations:
\begin{eqnarray}
  & 1. &  {\tt i=0}, ~~\source_i = \vec 0,~~ \Delta \vec \psi_i = \vec 0 \nonumber\\
  &    &  {\tt \vec \psi_0 = initial~model} \nonumber\\
  & 2. &  {\tt Determine~ D_s(\vec s_i), L_c(\vec \psi_0 + \Delta \vec \psi_i) } \nonumber\\
  & 3. &  {\tt Solve~ \left(M_c^T \,M_c + R\right) \vd = M_c^T \data}\nonumber\\
  &    &  {\tt ~~using~regularisation~parameters~\lambda_j}\nonumber\\
  & 4. &  {\tt Extract~\vec s_{i+1}~and~\delta \vec \psi~from~\vec r }\nonumber \\
  &    &  {\tt ~~\Delta \vec \psi_{i+1} = \Delta \vec \psi_{i} + \delta \vec \psi}\nonumber \\
  &    &  {\tt ~~i=i+1}\nonumber\\
  & 5. &  {\tt Has~G(\source_{i}, \vec \psi_0 + \Delta \vec \psi_i)~converged?}\nonumber\\
  &    &  {\tt ~~Yes: Exit}\nonumber\\
  &    &  {\tt ~~No~: Goto~2.}\nonumber
\end{eqnarray}

\noindent We have implemented this algorithm in Python, using
available packages for solving sparse-matrix equations. The code
should be relatively easy to parallelize. There are a number of issues
that might be worth considering when implementing the algorithm:

\begin{itemize}

\item[{\bf (a)}] Not all of the lens-plane pixels map onto the finite
source plane, and vice versa. We therefore mask those pixels in the
lens (source) plane for which the corresponding positions in the
source (lens) plane fall outside their pre-defined grids. We make sure
to avoid boundary issues (e.g.\ ill-defined gradients at the edge of
the grid). Masking is properly accounted for in the number of degrees
of freedom. Even though formally correct, because the majority of
masked pixels have no significant brightness (i.e.\ the grids are
of course chosen to capture the image and source brightness
distributions), it has little to no impact on the final result. We do
not determine in each iteration which source pixel (and corresponding
image pixels) are multiply imaged, to check whether they are
overconstrained. Although in principle this is the most proper
procedure, regularisation ensures that even un- or underconstrained
pixels are not overfitted (see Figure~1).

\item[{\bf (b)}] A linear gradient or additive constant in the
potential has no physical meaning (it has no corresponding
convergence). In each iteration, we therefore require the solution of
$\psi$ to have $\int \psi \equiv 0$ and $\int \nabla \psi \equiv 0$.
This also avoids the source to unnecessarily ``wander around'' in the
source plane. A similar issue is that of the mass-sheet degeneracy:
$\psi' \leftrightarrow (1-\kappa_{\rm sh})\psi +\frac{1}{2}\kappa_{\rm
sh}\,\theta^2$, where $\theta$ is some angular distance to a fiducial
centre. No observable, other than the time-delays change. The algorithm
is, in principle, unable to choose between any of these solutions,
which all have the same penalty value.  Since the mass enclosed within
the critical curves is well-determined (Kochanek 1991) {\sl and} the
mass-sheet $\kappa_{\rm sh}= 0$ for our starting model, also the final
solution of $\psi$ will contain only a small mass-sheet
contribution. By adding non-lensing information about the potential
(e.g.\ stellar kinematics; Koopmans 2004), the mass-sheet degeneracy
can be broken.

\item[{\bf (c)}] The brightest regions of the source often show the
strongest curvature (extrema are by definition far from linear). This
poses difficulties in reconstructing the brightest regions of the
lensed images, since regularisation tries to minimize
curvature. However, since the brightest areas also have the highest
S/N ratio and are therefore well constrained purely through the
$\chi^2$ term in the penalty function, one can downweight the
regularisation term for the brighter pixels in the source solution,
if needed.

\end{itemize}

\end{document}